\documentclass[baaa]{baaa}

\usepackage[pdftex]{hyperref}
\usepackage{subfigure}
\usepackage{natbib}
\usepackage{helvet,soul}
\usepackage[font=small]{caption}

\contriblanguage{1}

\contribtype{1}

\thematicarea{6}

\title{Studying a hot molecular core embedded in a photodissociation region}

\titlerunning{Molecules in a hot molecular core}

\author{
N.C. Martinez\inst{1,2},
M.B. Areal\inst{1},
\&
S. Paron\inst{1}
}

\authorrunning{Martinez et al.}

\contact{ncmarsan@gmail.com}

\institute{
Instituto de Astronom\'ia y F\'isica del Espacio, CONICET--UBA, Argentina
\and
Facultad de Ciencias Exactas y Naturales, UBA, Argentina
}

\resumen{
En el primer cuadrante de nuestra galaxia, en $l=33.\!\!^\circ134$, $b=-0.\!\!^\circ091$, aparece una extensa región de 
fotodisociación generada por un complejo de regiones H\textsc{ii}.
Asociado a esta región se encuentra gran cantidad de gas molecular con grumos en su interior. Embebido en uno de ellos, se localiza un núcleo molecular caliente conocido como G33.133-mm3.
Utilizando datos del James Clerk Maxwell Telescope (resolución angular de $\sim$15\arcsec), se estudió el cociente de abundancia $^{13}$CO/C$^{18}$O hacia el grumo molecular mencionado, y su relación con la radiación ultravioleta. Luego, en una escala espacial más pequeña, utilizando datos del Atacama Large Millimeter Array (resolución angular de 0.7\arcsec) se caracterizó al núcleo molecular G33.133-mm3, el cual posee un tamaño de $\sim$2600 ua y es un sitio propicio para la formación de estrellas. En particular, se mencionan algunos puntos sobre su química en base a la emisión del radical nitrilo (CN) y de las moléculas más complejas CH$_{3}$OH, CH$_{3}$CN, CH$_{3}$OCHO, y CH$_{3}$CCH. 
}

\abstract{At the first Galactic quadrant, at  $l=33.\!\!^\circ134$, $b=-0.\!\!^\circ091$, an extended photodissociation region
generated by an H\textsc{ii} region complex lies. This region is related to abundant molecular gas, and 
particularly, a hot molecular core, known as G33.133-mm3, appears embedded in a molecular clump.
Using data from the James Clerk Maxwell Telescope with an angular resolution of about 15\arcsec, we 
studied the $^{13}$CO/C$^{18}$O abundance ratio towards the mentioned molecular clump and its relation 
with the ultraviolet radiation. At smaller spatial scales, using data from the Atacama Large Millimeter Array (angular resolution about 0.7\arcsec), the hot molecular core G33.133-mm3, that has a size of about 2600 au, and is an appropriate site to form stars, was characterized. In particular, some points about its chemistry are mentioned based on the emission of the cyanide or nitrile radical (CN) and others more complex molecules, such as CH$_{3}$OH, CH$_{3}$CN, CH$_{3}$OCHO, and 
CH$_{3}$CCH.}


\begin{document}

\maketitle

\section{Introduction}

Hot molecular cores (HMCs) are compact ($\leq $0.1 pc), dense ($10^5 - 10^8$ cm$^{-3}$), and massive ($\sim$100 M$_{\odot}$) molecular structures related to the earliest phases of 
high$-$mass star formation (e.g. \citealt{beu07,motte18}). HMCs are usually embedded in molecular clouds and filaments \citep{rathborne06,lu2018}, which in turn can be related to photodissociation regions (PDRs), when complex of H\textsc{ii} regions are in the vicinity. Additionally, HMCs are the richest reservoirs of complex organic molecules in the Galaxy, including key species for prebiotic processes \citep{beltran18}. 

\begin{figure}[h!]
\centering
\includegraphics[width=0.45\textwidth]{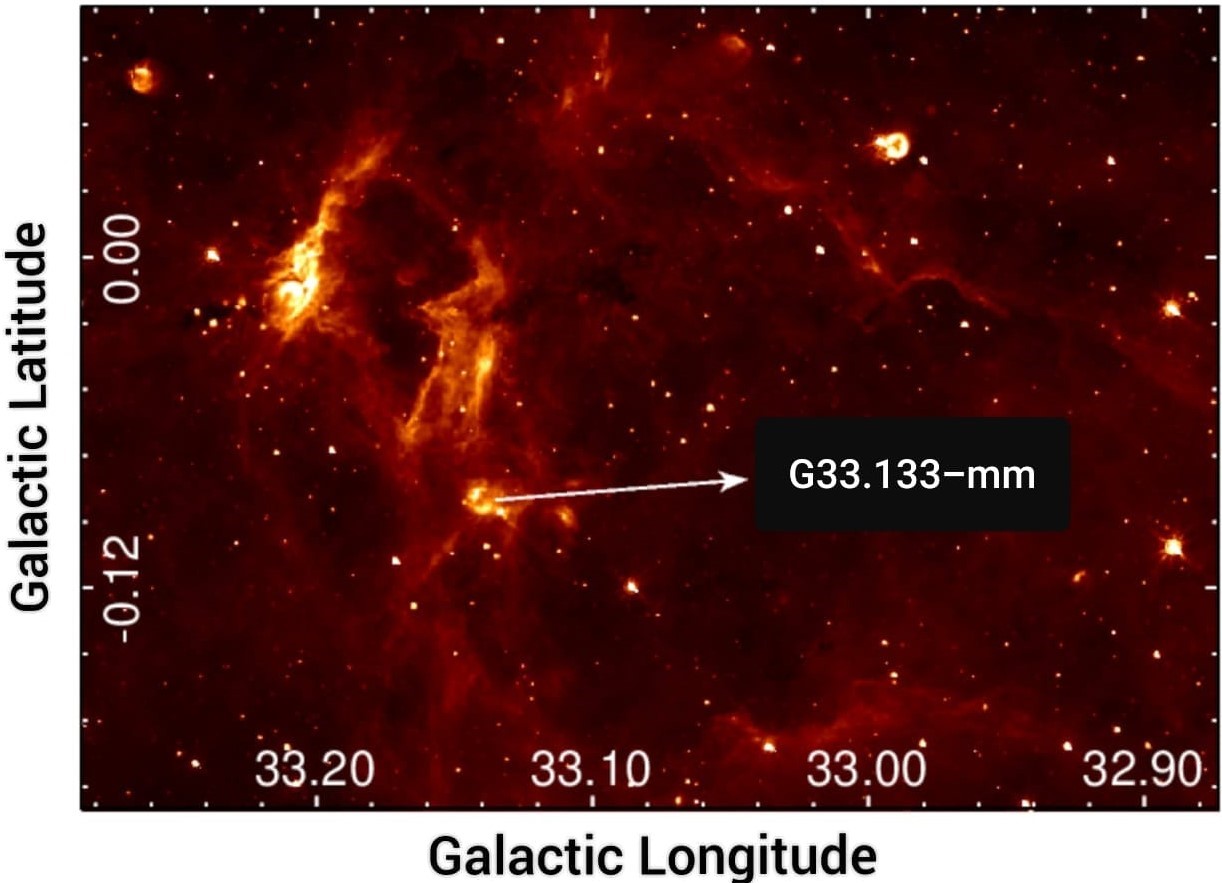} 
\caption{8 $\mu$m emission showing an extended photodissociation regions generated by a complex of H\textsc{ii} regions. The position of the HMCs complex (G33.133-mm) is indicated.}
\label{8um}
\end{figure}

\section{Presentation of the studied HMC} 

The HMC G33.133-mm3, with a size of about 2600 au \citep{paron21} is located at a distance of 4.5 kpc and 
it appears embedded in an extended PDR. G33.133-mm3 is part of a complex of at least four HMCs, suggesting that
fragmentation processes of a molecular clump are on going in the region. Figure\,\ref{8um} displays the 8 $\mu$m emission obtained
from the GLIMPSE/{\it Spitzer} survey\footnote{https://irsa.ipac.caltech.edu/data/SPITZER/GLIMPSE/} showing the PDRs and the position where the HMCs are located (named G33.133-mm). Given to the location of the HMCs, it is interesting to study the $^{13}$CO/C$^{18}$O abundance ratio, as done for instance by \citet{areal18} in many sources, because it gives us information about the relation between the UV radiation and the molecular gas at clump spatial scales. Additionally, we present a brief study regarding to the chemistry of G33.133-mm3 at core spatial scales.

\section{Data}

We used molecular data of the $^{12}$CO, $^{13}$CO and C$^{18}$O J=3--2 emission obtained from the public surveys generated by the 15-m James Clerk Maxwell Telescope: $^{12}$CO J=3--2 data was obtained from the COHRS survey \citep{2013ApJS..209....8D}, while the data of the other isotopes were extracted from the CHIMPS survey \citep{2016MNRAS.456.2885R}. The angular resolution of the data from each survey is 15\arcsec~and 14\arcsec, respectively.
The higher angular resolution data were obtained from the ALMA Science Archive\footnote{http://almascience.eso.org/aq/}. We used data from the project 2015.1.01312.S, that was observed in the ALMA Cycle 3 in configurations C36-2 and C36-3 in the 12~m array, at Band 6. We used the calibrated data which 
passed the second level of Quality Assurance (QA2).The angular resolution of this data set is about 0.7\arcsec, and the spectral resolution is 1.13 MHz.  

\section{Results}

\begin{figure}[h!]
\centering
\includegraphics[width=0.45\textwidth]{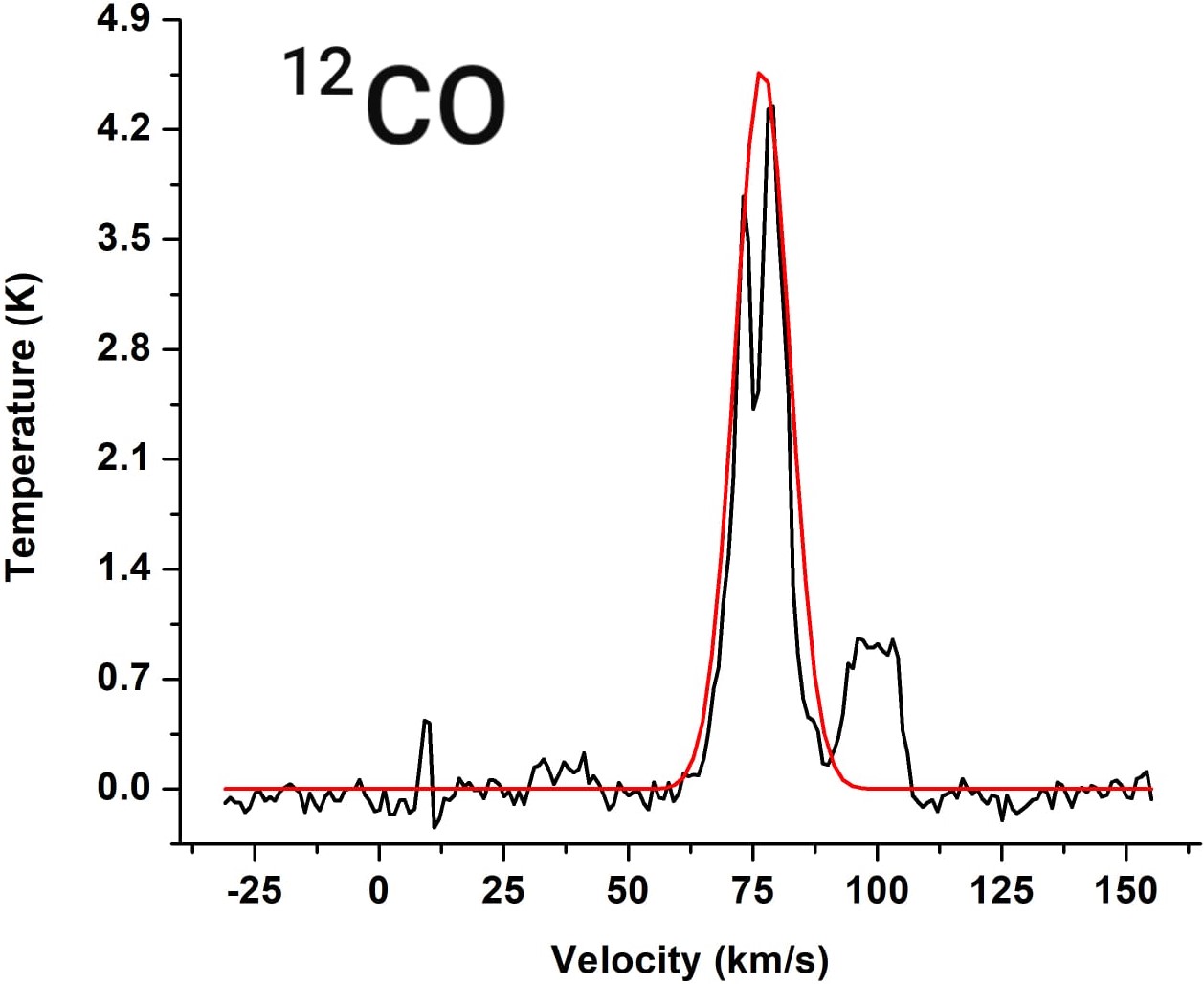}
\includegraphics[width=0.45\textwidth]{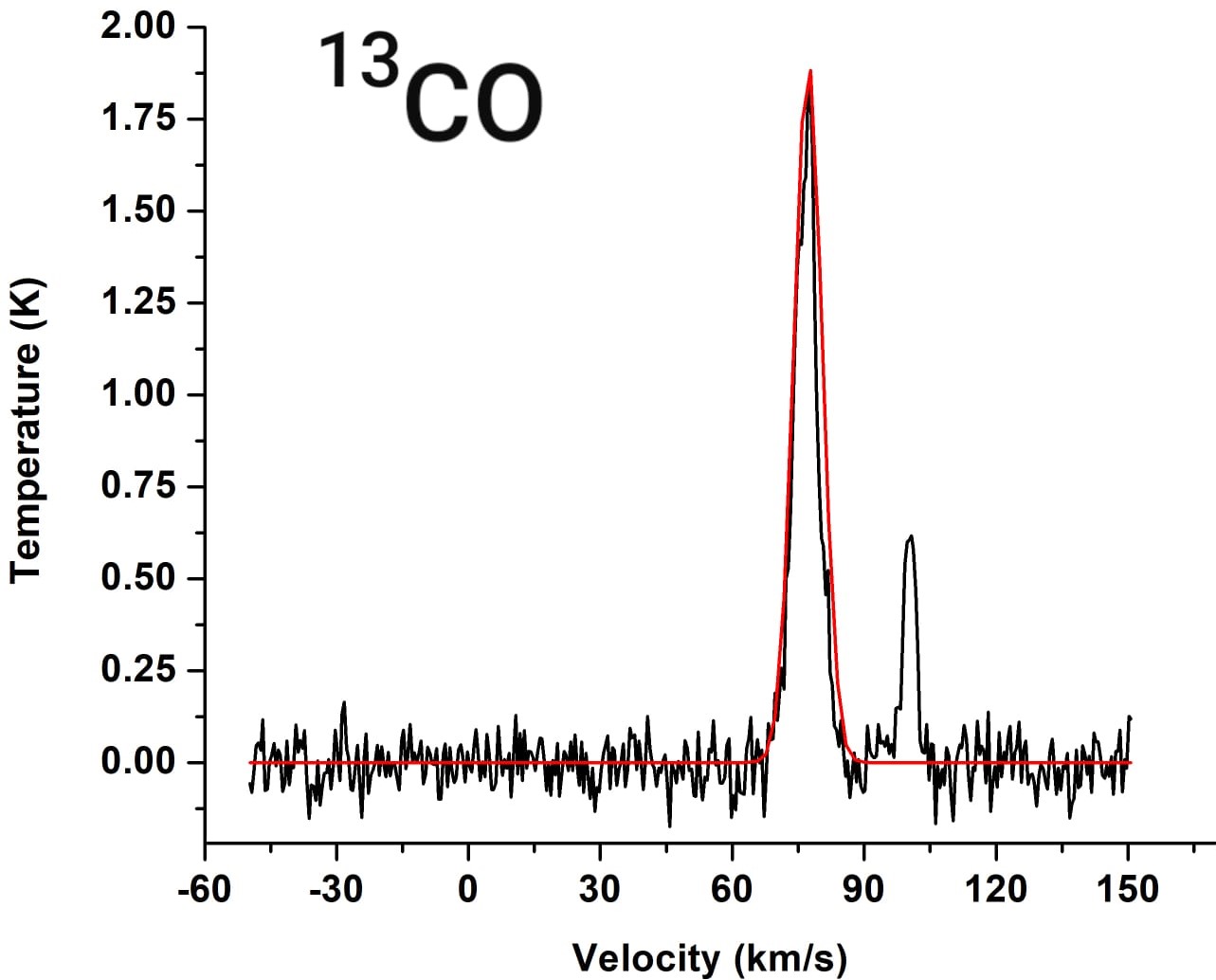}
\includegraphics[width=0.45\textwidth]{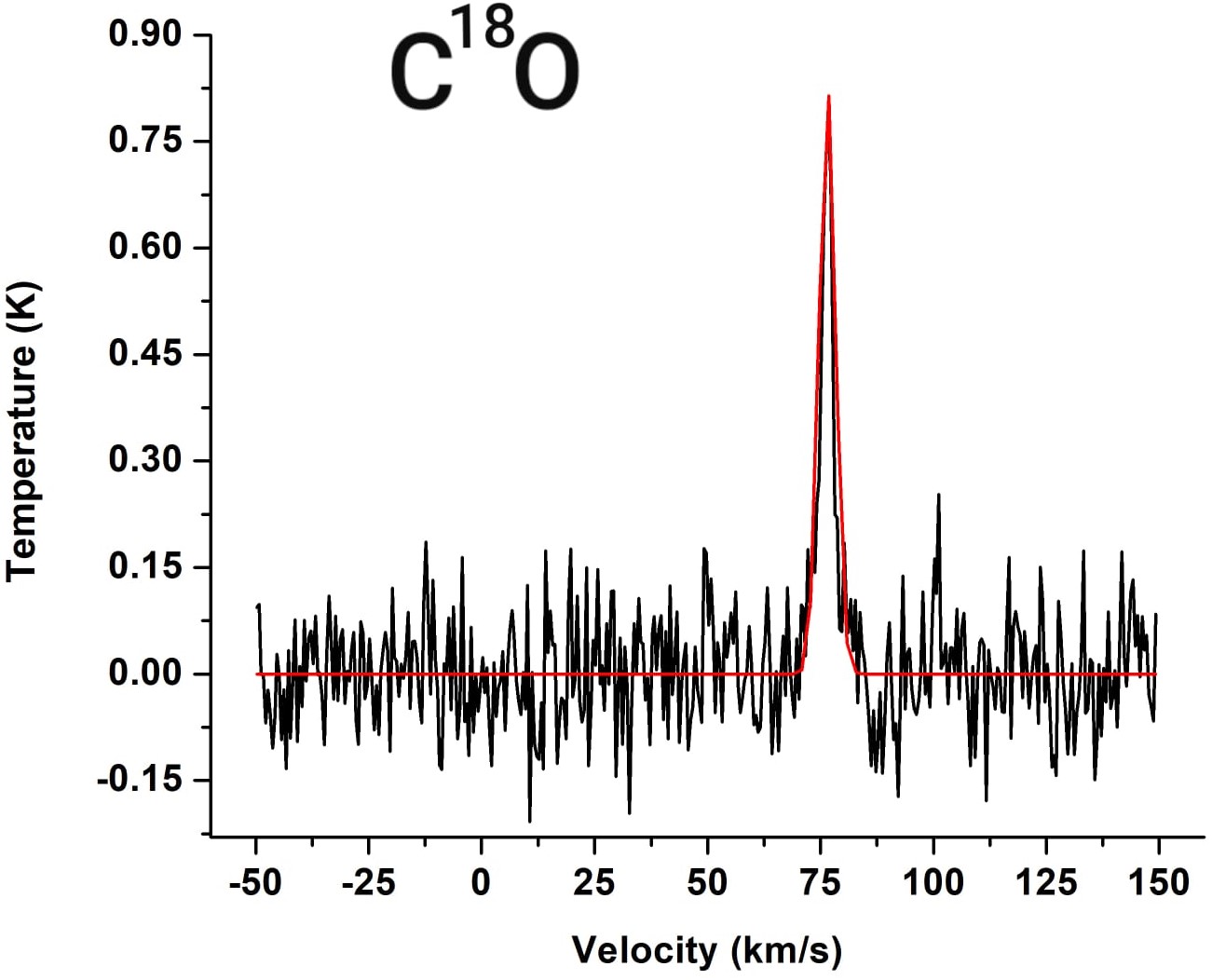}
\caption{Spectra of the CO isotopes obtained at the region where G33.133-mm3 is embedded. The red curves are Gaussian fitting to the 
molecular component at v$_{lsr} \sim 75$ km s$^{-1}$, the systemic velocity corresponding to the molecular clump in which the analyzed HMC is embedded. The obtained Gaussian parameters are not presented here for lack of space. In the case of the $^{12}$CO,
its emission appears self-absorbed, thus it was fitted by a single Gaussian that completes the line.}
\label{cospect}
\end{figure}

\subsection{The $^{13}$CO/C$^{18}$O abundance ratio}

Figure\,\ref{cospect} displays spectra of the CO isotopes obtained towards the molecular clump in which G33.133-mm3 is embedded. Given to the data beam size and considering the distance to the region, these spectra are probing molecular gas at a spatial scale of about 0.3 pc. By assuming local thermodynamic equilibrium (LTE), and following the typical formulae (see for instance \citealt{areal18}) we calculate the $^{13}$CO and C$^{18}$O column densities (N$^{13}$ and N$^{18}$) at the position of G33.133-mm3 to obtain the
$^{13}$CO/C$^{18}$O abundance ratio (X$^{13/18} =$ N$^{13}$/N$^{18}$). The obtained results,
including the excitation temperature ($T_{\rm ex}$) and the optical depths ($\tau$) are presented in Table\,\ref{cotable}. The obtained X$^{13/18}$ is compatible with values obtained at the borders of the densest regions within molecular clouds active in high-mass star formation \citep{paron18}. This ratio could be mapping the external gaseous layers of the dense clump in which G33.133-mm3 and others cores are embedded, where the ultraviolet radiation, responsible of the PDRs, is impinging.

\begin{table}
\centering
\caption{Parameters from the CO isotopes emission at the region where G33.133-mm3 is embedded.}
\label{cotable}
\begin{tabular}{lccccc}
\hline\hline
$T_{\rm ex}$     & $\tau^{13}$         &  $\tau^{18}$          &  N$^{13}$ &  N$^{18}$ &  X$^{13/18}$        \\
  (K)         &                     &                       &   (cm$^{-2}$) &     (cm$^{-2}$) &   \\       
\hline                 
  14          &   0.7               &   0.3                 &  1.67$\times10^{16} $ &    0.43$\times 10^{16}$  &          3.8 \\
\hline
\end{tabular}
\end{table}

\subsection{Complex molecules in G33.133-mm3}

\begin{figure}[h!]
\centering
\includegraphics[width=0.45\textwidth]{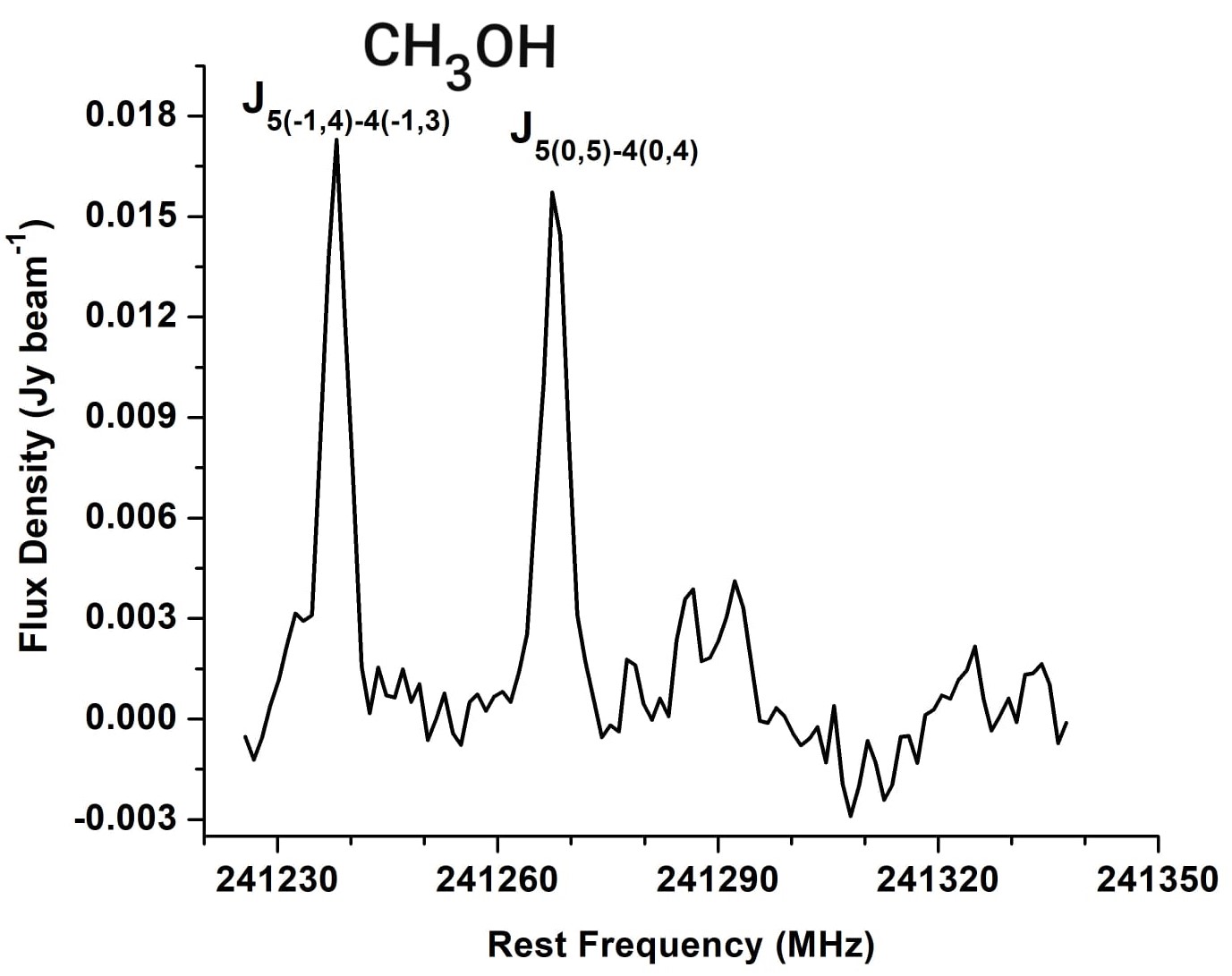}
\includegraphics[width=0.45\textwidth]{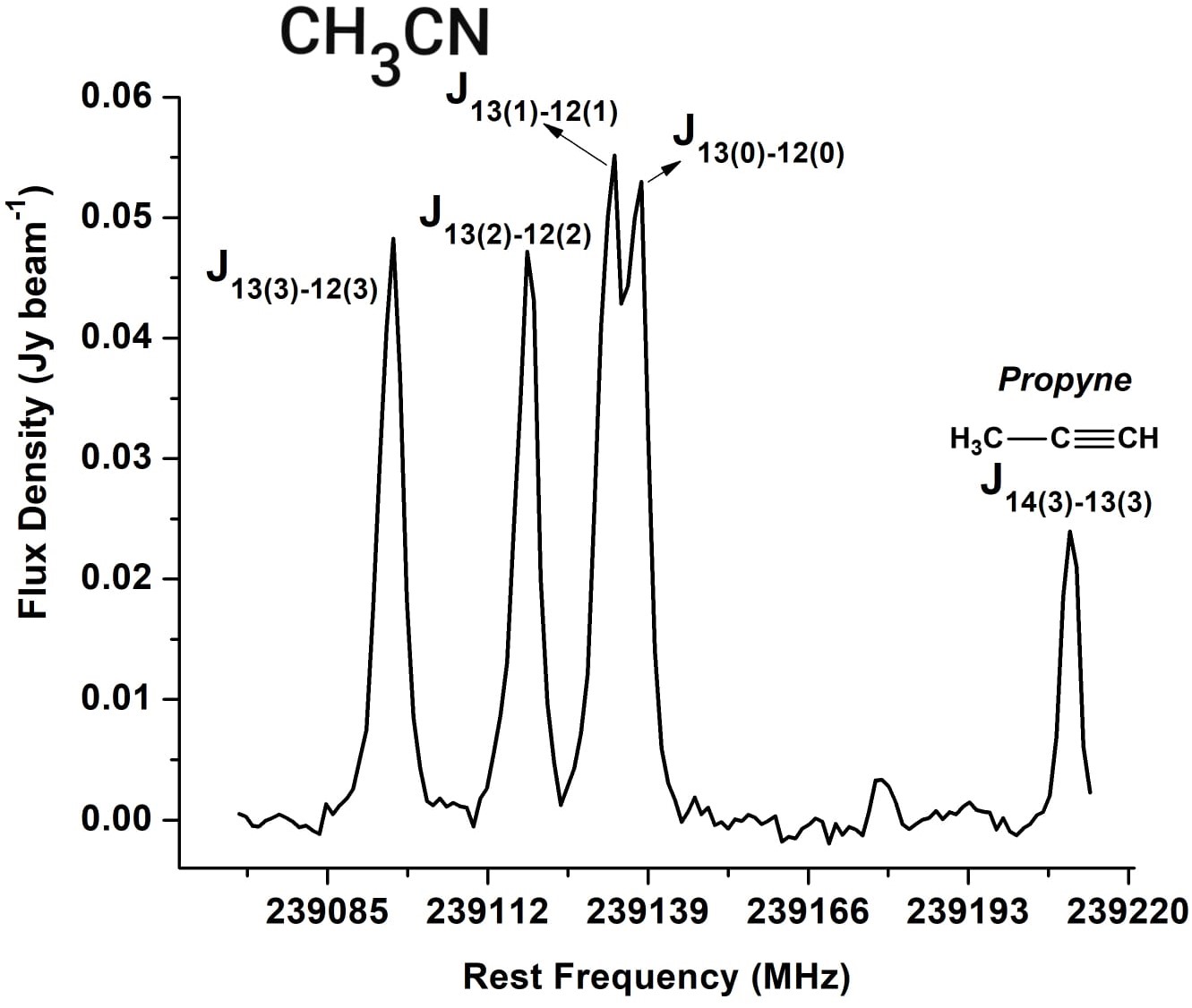}
\includegraphics[width=0.45\textwidth]{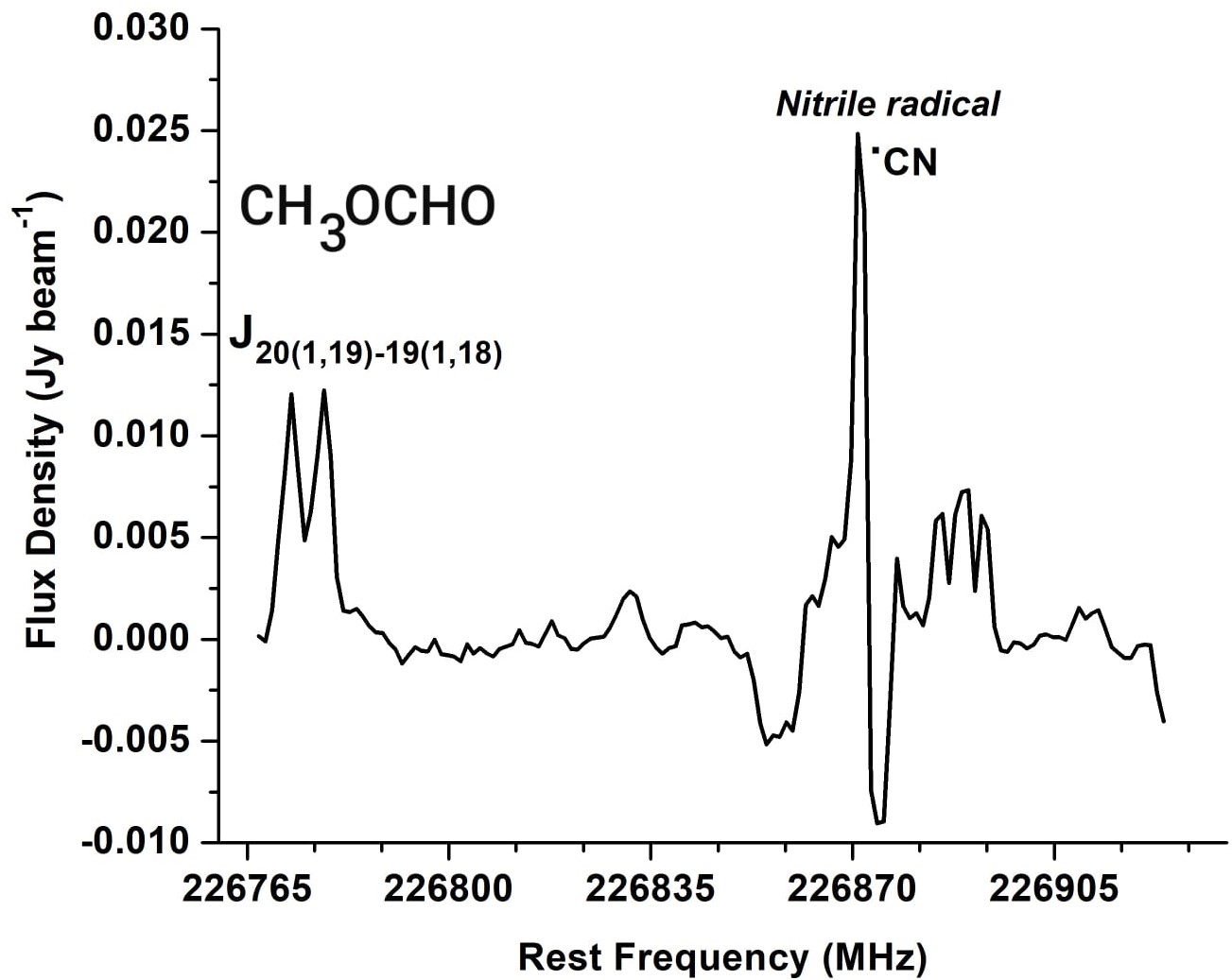}
\caption{Spectra of the complex molecules detected towards G33.133-mm3.}
\label{molecs}
\end{figure}

Figure\,\ref{molecs} displays spectra of the complex molecules detected towards the HMC G33.133-mm3 from the ALMA data. These spectra are probing gas at a spatial scale of about 0.02 pc. After converting 
the observed frequency to rest frequency, and following the NIST database\footnote{https://physics.nist.gov/cgi-bin/micro/table5/start.pl}, molecular species and their transitions were determined. They are indicated in the Figure 3.

From the ALMA data set, molecules such as methanol (CH$_{3}$OH), methyl formate (CH$_{3}$OCHO), and acetonitrile (CH$_{3}$CN) were detected. Additionally, in the spectrum of CH$_{3}$CN, emission of propyne (CH$_{3}$CCH) is also observed, while in the CH$_{3}$OCHO spectrum, it is also observed emission from the nitrile or cyanide radical (CN$^{.}$). The emission of these molecular species confirms the presence of high density gas, which is in agreement with the X$^{13/18}$ value discussed above. 
The observation of CN$^{.}$ at G33.133-mm3 encouraged the analysis of this radical towards a sample of HMCs presented in \citet{paron21}, in which it was concluded that the presence of CN$^{.}$ seems to be ubiquitous along the different star formation stages.

It is known that CH$_{3}$CN, CH$_{3}$OCHO, and CH$_{3}$OH are well tracers of hot molecular cores/corinos (e.g. \citealt{areal20,molet19, beltran18}). These complex molecular species form in the dust grain surfaces, and when the temperature increases, they thermally 
desorbe from the dust. Particularly, when the temperature of a molecular core reaches about 90 K, CH$_{3}$OH thermally desorbes from the grain mantles, and its gas-phase abundance is enhanced close to the protostars \citep{brown07}. In the case of propyne, this molecular species is formed on the 
grain surface through successive hydrogenation of physisorbed C$_{3}$ \citep{hick16}, and it is an important molecule for large comparison studies of chemical diversity among star-forming regions \citep{tani18}.

G33.133-mm3 is part of a large sample of HCMs whose chemistry and their relation with star-forming processes will be studied in future works by this group.

\begin{acknowledgement}

This work was partially supported by grants PICT 2015-1759 - ANPCYT, and
PIP 2021 11220200100012 - CONICET. The authors, specially N.C.M., 
thank to the scientific committee of the Asociación Argentina de Astronomía (AAA) Annual Meeting for
the mention given to the poster presented during the meeting. These results are part of the work done
by N.C.M. during the 2021 summer AAA fellowship program.

\end{acknowledgement}


\bibliographystyle{baaa}
\small
\bibliography{bibliografia}

\begin{thebibliography}{15}
\providecommand{\natexlab}[1]{#1}

\bibitem[{{Areal} et~al.(2018)}]{areal18}
{Areal} M.B., et~al., 2018, \aap, 612, A117

\bibitem[{{Areal} et~al.(2020)}]{areal20}
{Areal} M.B., et~al., 2020, \aap, 641, A104

\bibitem[{{Beltr{\'a}n} \& {Rivilla}(2018)}]{beltran18}
{Beltr{\'a}n} M.T., {Rivilla} V.M., 2018, E.~{Murphy} (Ed.), \textit{Science
  with a Next Generation Very Large Array}, \textit{Astronomical Society of the
  Pacific Conference Series}, vol. 517, 249

\bibitem[{{Beuther} et~al.(2007)}]{beu07}
{Beuther} H., et~al., 2007, B.~{Reipurth}, D.~{Jewitt}, K.~{Keil} (Eds.),
  \textit{Protostars and Planets V}, 165

\bibitem[{{Brown} \& {Bolina}(2007)}]{brown07}
{Brown} W.A., {Bolina} A.S., 2007, \mnras, 374, 1006

\bibitem[{{Dempsey} et~al.(2013){Dempsey}, {Thomas} \&
  {Currie}}]{2013ApJS..209....8D}
{Dempsey} J.T., {Thomas} H.S., {Currie} M.J., 2013, \apjs, 209, 8

\bibitem[{{Hickson} et~al.(2016){Hickson}, {Wakelam} \& {Loison}}]{hick16}
{Hickson} K.M., {Wakelam} V., {Loison} J.C., 2016, Molecular Astrophysics, 3, 1

\bibitem[{{Lu} et~al.(2018)}]{lu2018}
{Lu} X., et~al., 2018, \apj, 855, 9

\bibitem[{{Molet} et~al.(2019)}]{molet19}
{Molet} J., et~al., 2019, \aap, 626, A132

\bibitem[{{Motte} et~al.(2018){Motte}, {Bontemps} \& {Louvet}}]{motte18}
{Motte} F., {Bontemps} S., {Louvet} F., 2018, \araa, 56, 41

\bibitem[{{Paron} et~al.(2018){Paron}, {Areal} \& {Ortega}}]{paron18}
{Paron} S., {Areal} M.B., {Ortega} M.E., 2018, \aap, 617, A14

\bibitem[{{Paron} et~al.(2021)}]{paron21}
{Paron} S., et~al., 2021, \aap, 653, A77

\bibitem[{{Rathborne} et~al.(2006){Rathborne}, {Jackson} \&
  {Simon}}]{rathborne06}
{Rathborne} J.M., {Jackson} J.M., {Simon} R., 2006, \apj, 641, 389

\bibitem[{{Rigby} et~al.(2016)}]{2016MNRAS.456.2885R}
{Rigby} A.J., et~al., 2016, \mnras, 456, 2885

\bibitem[{{Taniguchi} et~al.(2018)}]{tani18}
{Taniguchi} K., et~al., 2018, \apj, 866, 150

\end{thebibliography}
 
\end{document}